# TWO GENERATIONS IN STELLAR COMPLEXES AND ASSOCIATIONS IN M 33 GALAXY AND THEIR SPATIAL CORRELATION

TODOR VELTCHEV, NINA KOLEVA, PETKO NEDIALKOV
and GEORGI R. IVANOV

*Department of Astronomy, Faculty of Physics, Sofia University,
5 James Bourchier Blvd., 1164 Sofia, Bulgaria*
*e-mail: eirene@phys.uni-sofia.bg*

**Abstract.** Massive stellar content of stellar complexes and associations in M 33 is studied combining deep UBV photometry from the Local Group Survey (Massey et al. 2006) and JHK photometry from the 2MASS. Two basic populations (incl. OB stars and red supergiants) are distinguished and their application for reconstruction of the star formation process in this galaxy are discussed.

## 1. INTRODUCTION

Photometric studies of the massive stellar content in nearby galaxies have gained a new impact in the last decade due to the advance of observational astronomy. Special attention attract compact groups of massive stars (complexes, associations, young clusters) since they trace star formation at different scales. The detailed investigation of such groups and especially of their crowded parts faces at least two problems. Some very luminous "stars" in the group turn out to be blended images of several stars which causes anomalous colors and widenning of the Main sequence (MS) on CMD (e.g. Haiman et al. 1994). On the other side, the low photometric limit of observational data taken with ground-based telescopes with apertures about and below 2m does not allow a reconstruction of substantial parts of the MS and the branch of the Red supergiants (RSGs). Coverage of the disks of nearby galaxies by frames obtained with HST (WFPC2, ACS) is still not sufficient for an extensive study of star formation.

The recently published Local Group Survey (LGS), based on CCD imaging on KPNO and CTIO 4m- telescopes (Massey et al. 2006), resolves or, at least, alleviates the mentioned problems. Its precise UBVRI photometry allows reliable selection of blue and red massive stars in a nearby galaxy and subsequent analysis of color-magnitude (CMD) and color-color diagrams. Additionally, one may use the 2MASS catalog as a source for possible RSG candidates since its infrared stellar photometry gives opportunity to distinguish such objects from foreground dwarfs.

This paper presents further steps in our study of the massive stellar content in star formation sites in the Triangulum galaxy (M 33). This study started with dereddening of OB stars and age estimation in 17 classical associations (Koleva et al. 2006) and is now enlarged with inclusion of RSG candidates.

## 2. STELLAR PHOTOMETRY AND ITS DEREDDENING

### 2.1. OB stars

The initial sample includes 28 378 blue massive stars with reliable UBVRI photometry from the LGS survey addressing selection criterion (B-V)<1., i.e. within large range of reddening. Successfull dereddening was performed by Vassileva et al. (2006) for ≈15000 stars through the classical Q-method on diagram (B-V) vs. (U-V), using zero-absorption main sequence from Bessell (1990). The procedure removes most of the blended images (that lie above the reddening line, corresponding to bluest possible color $(U-V)_0$).

### 2.2. RSG candidates

We selected initially 806 stars with JHK photometry from 2MASS and identified with sources from LGS that fall within the boundaries of the 'classical' OB associations or of some of their subgroups, as outlined by Humphreys and Sandage (1980) and, respectively, Ivanov (1987, 1991). Only 96 of them seem to be RSG candidates judging on their optical colors ((B-V)>1.7). After additional expection of the IR colors some of those objects turn out to be severely reddened OB stars. Eventually, the typical detected RSG membership (if any) of the rich OB associations is 2-4 stars. They allow dereddening through the Q-method on diagram (H-K) vs. (J-K) using interstellar extinction law according to Rieke and Lebofsky (1985). As seen in Fig. 1, the group of foreground stars ((J-K)<0.6) is clearly separated from that of the RSG candidates.

## 3. STELLAR GROUPS

The locations of the selected OB stars and the RSG candidates in the field of M 33 are shown in Fig. 2. The transformation from equatorial to rectangular XY coordinates was done adopting a positional angle PA=23°. We have delineated 5 conventional fields ("Center", "North", "South", "East" and "West") that cover the OB associations or complexes containing at least one RSG candidate. The central field encompasses the very populated circumuclear region and - in particular, - the frame studied by Wilson (1991) by use of UBV CCD photometric observations with the 3.6-m CFHT and the 60-in Palomar telescope. This is the region where the most pronounced spiral arms SI and NI begin. The stellar density in the other 4 fields is significantly lower. Fields "East" and "West" cover the ends of the main arms while "North" and "South" - the secondary arms.

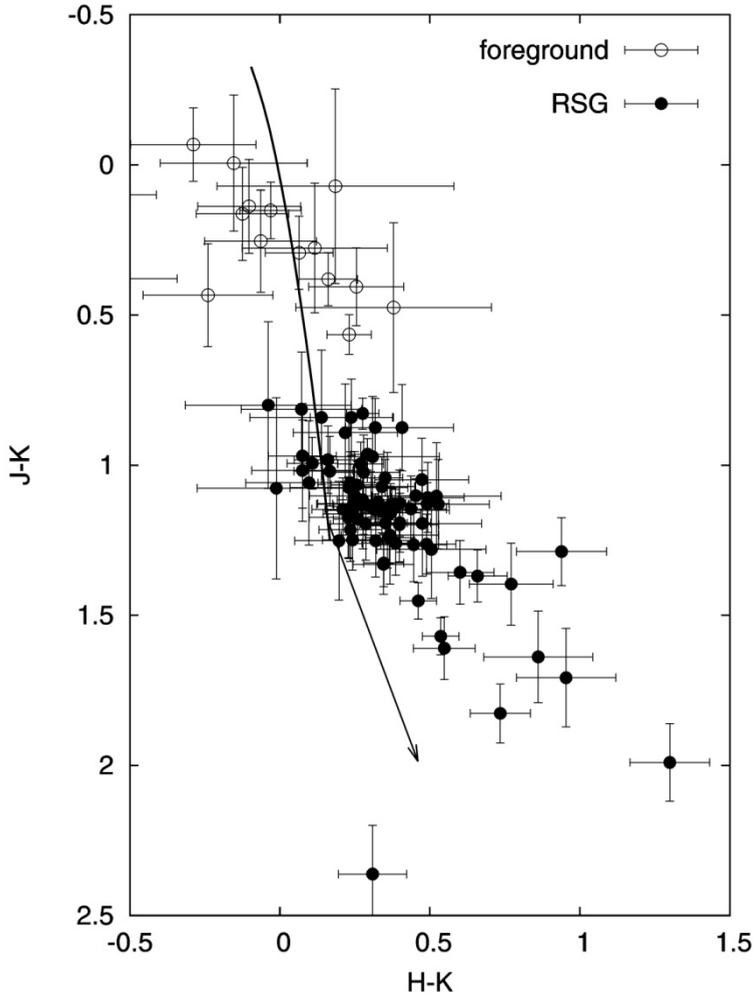

**Figure 1:** Two-color diagram of the RSG candidates within the classical OB associations. The reddening vector and the approximation of the zero-absorption line are plotted.

We report here first results towards discrimination of two main generations in 8 classical associations. The latter are located in the fields "Center" and "West" and contain 1 to 3 identified RSGs, confirmed through their true optical and IR colors (Fig. 3). Associations Nr. 50, 53, 96 and 137 are divided into 3 subgroups (denoted a, b, c).

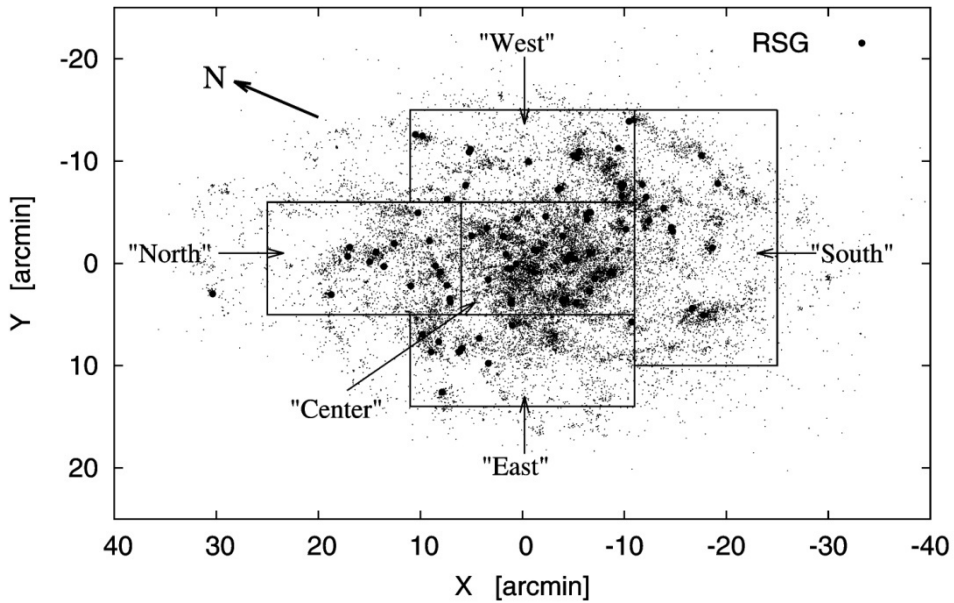

**Figure 2:** The dereddened OB stars in M 33 and locations of the RSG candidates (large dots). The oultines of chosen 5 fields are shown.

## 4. CM DIAGRAMS

The selected dereddened OB stars within the outlines of the associations were plotted on CMDs $(B-V)_0$ vs. $M_V$ while the RSGs candidates were plotted on CMDs $(J-K)_0$ vs. $M_K$. Several such plots are given in Fig. 4. The samples of blue stars were restricted additionally through criterion $(B-V)<0.4$ in order to minimize possible foreground contamination. The isochrones of the Padova group for $Z=0.008$ (Bertelli et al. 1994) give reliable estimates for the ages of generations of OB stars due to small photometric errors of the samples. Such estimates in the RSG case are rather rough but may serve as a consistency check. Brief inspection of Table~1 shows a good consistency between the age estimations of the older generation using OB stars and using RSGs. The existence of at least two distinct generations within a 'classical association' points to a picture of ongoing and, possibly, discrete process of star formation within spatially restricted area. Thus the next natural step is to investigate do they exhibit any specific internal structure of the given association.

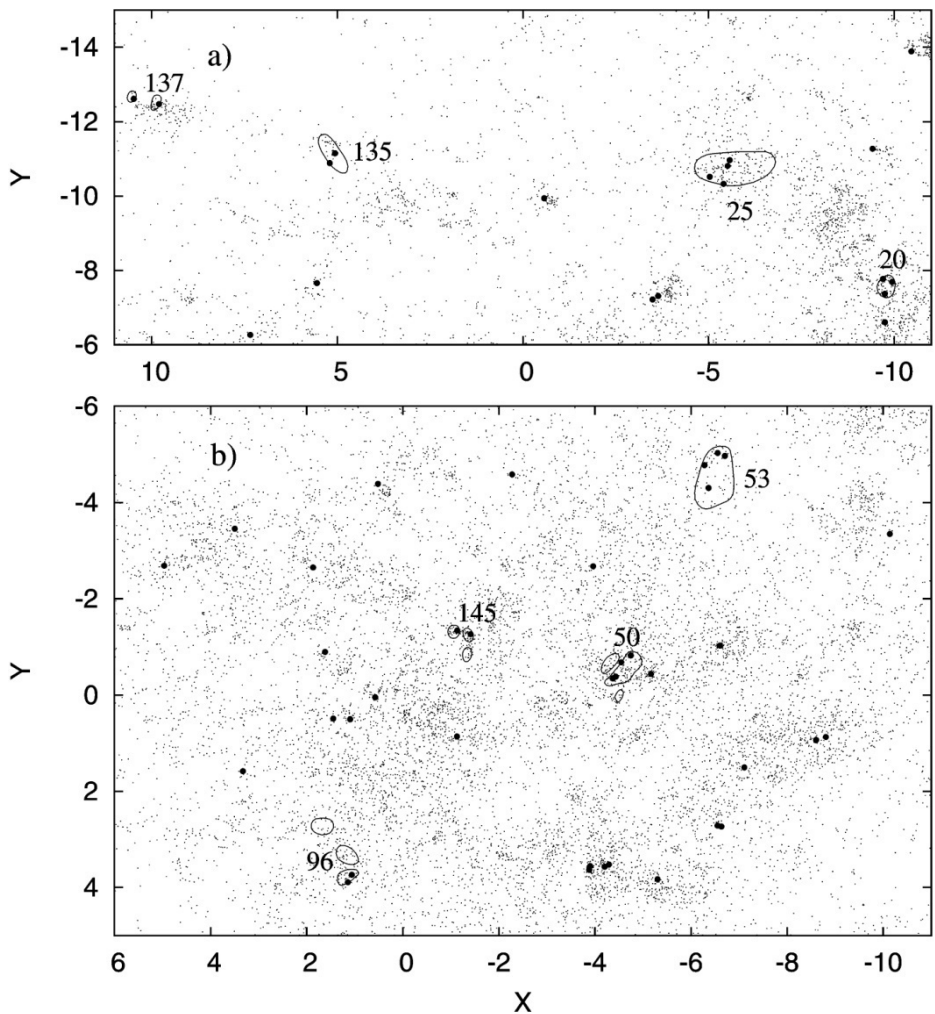

**Figure 3:** Eight studied associations in the fields "West" (a) and "Center" (b). The locations of the RSGs are plotted with large dots.

## 5. SPATIAL DISTRIBUTION

The 2D distribution of both generations (provisionally called 'Generation 1' and 'Generation 2') on the XY plane and within some of the associations is shown in Fig. 5. Generation 1 is more clustered and often clumpy while the older one is rather diffuse and shows no specific structure. The generations are seemingly segregated and occupy different spatial areas. For a quantitative analysis of their spatial correlation we use the technique, presented in Ivanov (1998). It is based on supposition for a Poissonian distribution of each of the two populations and gives the percentage of N5 associated couples between them: R5=N5/N, where N is the total number of couples. The associated couples are selected by criterion $P_{12}(k)<0.05$, where $P_{12}(k)$ is the probability to find two stars of both populations within radius $d_k$, the angular distance between the stars of the $k^{th}$ stellar couple. On the other side, probability $P_{12}(k)>0.95$ is a plausible measure for mutual seggregation of the populations in question - this criterion defines the 'foreground' or not associated couples ($N_{fgr}$). The number ratio RN5=N5/$N_{fgr}$ is a correlation parameter.

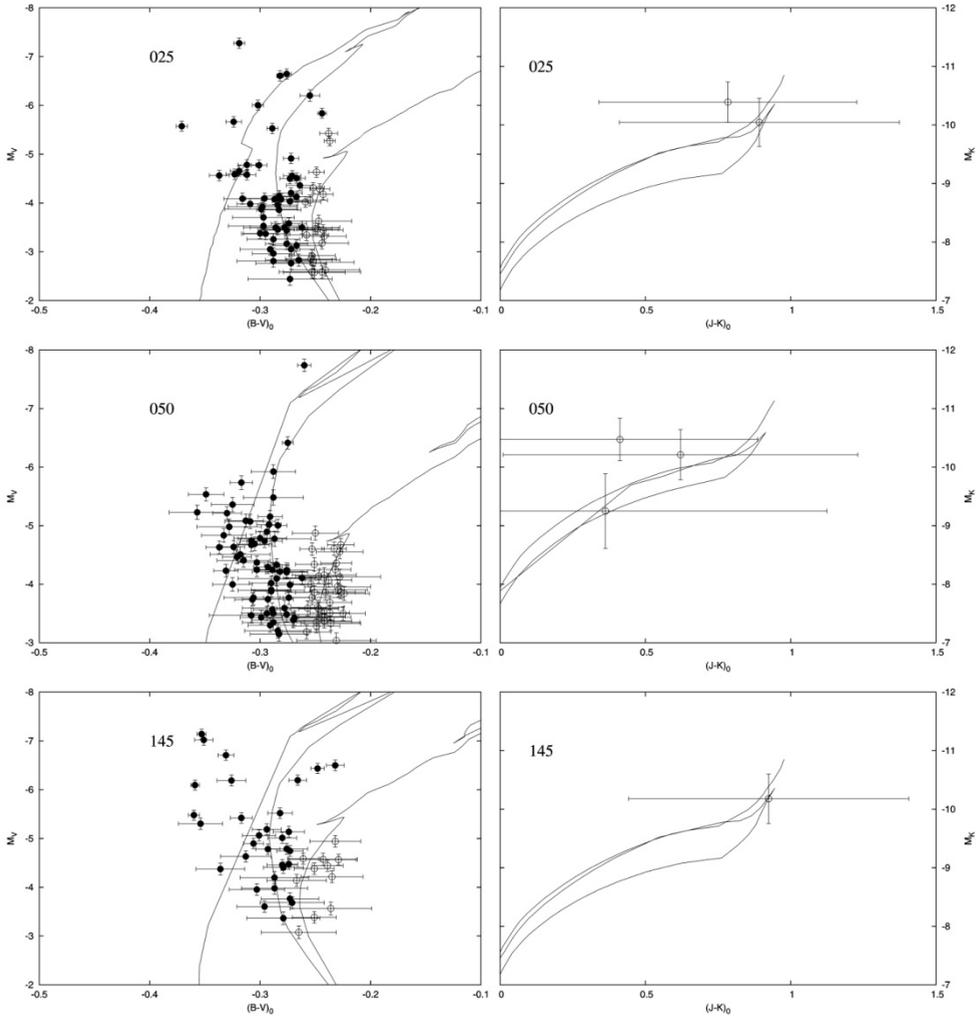

**Figure 4:** Color-magnitude diagrams for the selected OB stars (left) and RSGs (right) in the associations Nr. 25, 50 and 145. The ages of the isochrones are given in Table 1.

The quantities R5 and RN5 for the studied associations are specified in the last two columns of Table 1. Their values (<1) confirm the lack of spatial correlation between Generation 1 and Generation 2. The age difference of $\approx 10^7$ yr between the generations corresponds roughly to the lifetime of stars with $\gtrsim 20$ $M_\odot$ (e.g. Bertelli et al. 1994). Taking in view also the clumpy structure of the youngest stars and the spatial distribution of their older peers, that lends support to the hypothesis that SN bursts of the most massive population are the main driver (through shock waves) of star formation within an association.

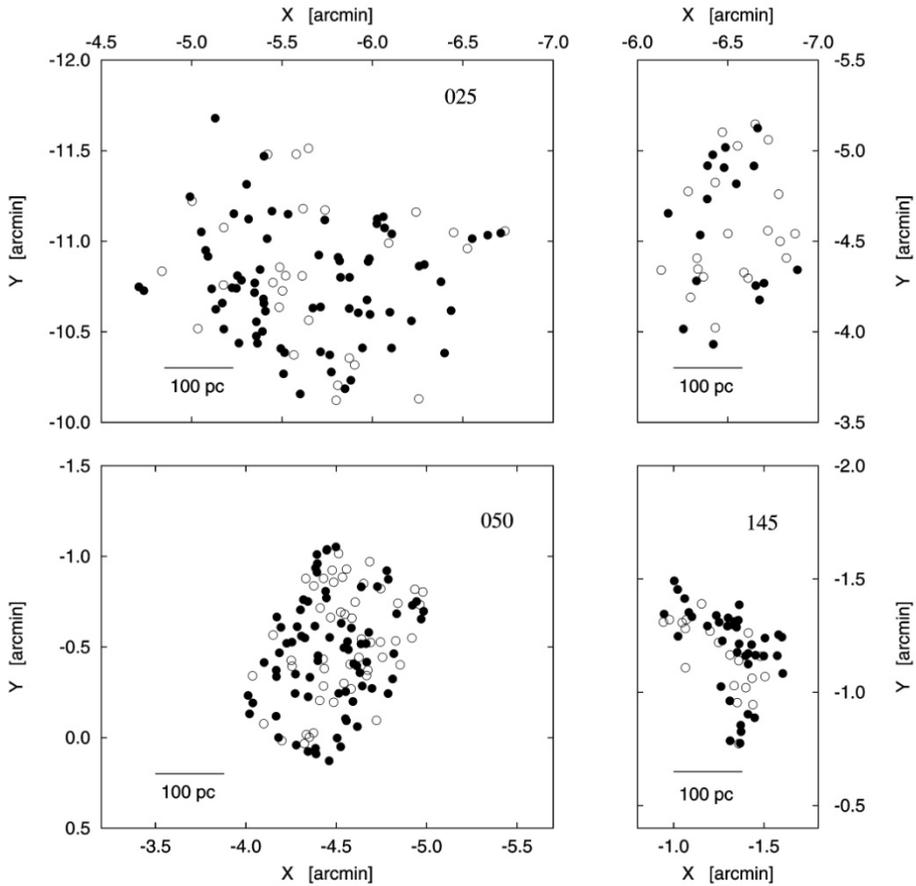

**Figure 5:** Spatial distribution of the youngest Generation 1 (filled symbols) and the older Generation 2 (open symbols) in 4 of the studied associations.

**Table 1.** The two generations in the chosen 'classical associations' and their correlation. Number of objects and age estimations (in units lg t), obtained through inspection of the CMDs, are given. Typical age dispersion/uncertainty is Δ(lg t)≈ 0.1 for the OB stars and Δ(lg t) ≈ 0.2 for the RSGs (see Fig. 4). The values of the parameters R5 and RN5 point to a lack of spatial correlation between the generations (see text).

| Assoc. # | Generation 1 OB stars | Age | Generation 2 OB | RSGs | Age OB | RSGs | R5 | N5 |
|---|---|---|---|---|---|---|---|---|
| 020 | 28 | 6.70 | 7 | 2 | 7.15 | 7.15 | 0.00 | 0.00 |
| 025 | 73 | 6.70 | 26 | 2 | 7.10 | 7.10 | 0.04 | 0.17 |
| 050 | 76 | 6.60 | 47 | 3 | 7.10 | 7.00 | 0.04 | 0.08 |
| 053 | 17 | 6.60 | 17 | 3 | 7.20 | 7.20 | 0.18 | 0.75 |
| 096 | 23 | 6.60 | 15 | 2 | 7.30 | 7.40 | 0.13 | 0.40 |
| 135 | 7 | 6.60 | 7 | 1 | 7.05 | 7.05 | - | - |
| 137 | 39 | 6.60 | 5 | 1 | 7.10 | 7.10 | - | - |
| 145 | 39 | 6.60 | 19 | 1 | 7.10 | 7.10 | 0.15 | 0.60 |

## Acknowledgement

This research was partially supported by contract Nr. F-201/06 with Scientific Researh Foundation, Ministry of Education and Sciences, Bulgaria.